\documentclass[10pt]{article}
\usepackage{amssymb,amsmath,latexsym,graphicx,hyperref,float,fullpage,enumitem}
\usepackage[font=footnotesize,labelfont=bf, justification=justified]{caption}
\usepackage{subfigure}
\hypersetup{colorlinks,urlcolor=magenta,linkcolor=blue,citecolor=black}
\usepackage{fancyhdr}
\usepackage{amsmath}
\usepackage{fullpage}
\usepackage{array}
\usepackage{graphicx}
\usepackage{mathrsfs}
\usepackage{amsthm}

\usepackage{enumitem}
\usepackage{amsfonts}
\usepackage[superscript,biblabel]{cite}
\usepackage[parfill]{parskip}
\usepackage{authblk}
\usepackage{booktabs}
\usepackage{rotating}

\begin{document}
\title{Voter Models and External Influence}
\author{Jimit Majmudar\thanks{\textit{Department of Mathematics, University of British Columbia Okanagan,} e-mail: jimit@alumni.ubc.ca} , Stephen M. Krone\thanks{\textit{Department of Mathematics, University of Idaho}} , Bert O. Baumgaertner\thanks{
\textit{Department of Philosophy, University of Idaho}} , Rebecca C. Tyson\thanks{\textit{Department of Mathematics, University of British Columbia Okanagan}}}

\date{\today}
\maketitle

\begin{abstract}
In this paper, we extend the voter model (VM) and the threshold voter model (TVM) to include external influences modelled as a jump process. We study the newly-formulated models both analytically and computationally, 
employing diffusion approximations and mean field approximations. We derive results pertaining to the probability of reaching consensus on a particular opinion and 
also the expected consensus time. We find that although including an external influence leads to a faster consensus in general, this effect is more pronounced in the VM as compared to the TVM.
Our findings suggest the potential importance of ``macro-level'' phenomena such as the external influences as compared to ``micro-level'' local interactions. \newline

\end{abstract}

\section{Introduction}
\label{intro}
Human opinions and collective human behaviours have been studied for more than a century \cite{mackay, le1897, sherif1936, asch1956,milgram1974,keynes2006,schelling1971}. Relatively recent is the mathematical analyses of these dynamics, which 
have been made possible due to seminal frameworks such as the voter model (VM) \cite{Clifford1973, HolleyRichardandLiggett1975}, DeGroot learning \cite{DeGroot1974}, and the naming game model \cite{Baronchelli2005, DallAsta2006}. 
A significant number of advances in opinion dynamics have their roots in statistical physics \cite{nyczka2012, vazquez2008, ben1996}. For details on the origins and evolution of this domain, the reader is referred 
to the comprehensive review by Castellano et al \cite{castellano2009}.

While refining the mechanistic description of node-to-node interactions has received considerable attention \cite{Cox1991, VanZandt, Hegselmann2002}, the incorporation of external influences  
has received little modelling scrutiny, one reason for which seems to be the associated loss of analytical tractability. There is evidence, however, that media can play an important role in opinion dynamics in many different contexts, for example, 
in climate change \cite{Brulle2012}, and in electoral voting \cite{druckman2005, beck2002}. 
Furthermore, commercial relevance of this phenomenon can be found in quantitative marketing, when multiple brands compete
to sell their respective products via advertising. A theoretical examination of the effect of external influence is thus important. Existing work, however, is either not amenable to detailed mathematical analysis \cite{Quattrociocchi2014}, or lacks 
generality \cite{Fotouhi2013}. We address these shortcomings by modelling external influence as a jump process, and use the theory of jump diffusion processes. Jump diffusion models are widely used 
in financial domains such as derivative pricing and risk management \cite{Kou2007}. 

The organisation of the paper is as follows: In section \ref{models} we define our models, followed by our results and their interpretation in section \ref{results}. We finish with a high-level discussion in section \ref{disc}.

\section{Models}
\label{models}
Let $G$ be a $k-$regular undirected graph of $N$ nodes, where each node is considered to be an individual agent and the links represent social connections between the agents. Let $S(G)$ denote the set of nodes of $G$.
Consider a binary opinion space, i.e., a node $x \in S(G)$ must have either opinion $0$ or opinion $1$ at time $t$ which is denoted by $s_x(t)$. Let $A$ be the adjacency matrix, with the element $A_{xy}$ 
being $1$ if nodes $x$ and $y$ are connected, and $0$ otherwise. The neighbourhood of a node consists of the immediate nodes with which it is connected. \textit{Consensus} is said to be reached when either all nodes have opinion $0$ or opinion $1$, and 
it is assumed that these states are absorbing. Consensus on opinion $1$ is herein called \textit{fixation}.

  \subsection{Jump Voter Model}
  The jump voter model (JVM) is a discrete-time process that is updated according to two rules. At each time step, one of the following occurs: \newline
  \begin{enumerate}
   \item With probability $(1-p)$, a single node is randomly selected which then adopts the opinion of one of its neighbours chosen randomly. \newline
   \item With probability $p$, either a random number of $0$ opinion nodes update their opinions to $1$ or a random number of $1$ opinion nodes update their opinions to $0$. If this number exceeds the number of nodes available to be updated, then the 
   all the available nodes are updated. A signed form of this random variable $(Z)$ follows the convention that it is negative in the case of the former update, and positive for the latter. \newline
  \end{enumerate}

  The first update rule captures the node-to-node interactions of the ``classical'' VM. The second rule captures 
  the more global external influence that makes several opinions flip simultaneously, a phenomenon that we call a jump. Call $p$ the 
  jump probability, and note that when $p=0$ the model reduces to a discrete-time version of the VM on a graph structure. The JVM is a discrete state space Markov chain, and the total 
  number of opinion $1$ nodes in the graph at time step $t$, denoted by $X^N(t)$, can be thought of as a global summary statistic of that Markov chain. More formally,
  \begin{equation}
  X^N(t) = \sum_{x \in S(G)} s_x(t). \nonumber
  \end{equation}
  
  The jump random variable $Z$ is independent of the state of the process, and we define the scaled jump $Y \equiv Z/N$. We restrict our attention to the case where the mean of $Y$ is zero, that is the case of no bias in the external 
  influence. The variance of $Y$, denoted by $\mathrm{v}$, represents the strength of the external influence, and we call it the jump variance. (A brief summary of all the model parameters is provided in Table \ref{pars}.) 

  \subsection{Jump Voter Diffusion}
  We make use of a diffusion approximation here and derive a jump-diffusion process to which the JVM weakly converges. To proceed with this jump diffusion approximation, we will need the transition probabilities corresponding 
  to update rule $1$ (i.e. $p=0$):
  
  \begin{itemize}
   \item $P[i \to i+1] \equiv$ probability that a $0 \to 1$ update happens at a certain time step when the count of nodes with opinion $1$ is $i$, and,
   \item $P[i \to i-1] \equiv$ probability that a $1 \to 0$ update happens at a certain time step when the count of nodes with opinion $1$ is $i$.
  \end{itemize}
  
  \iffalse
  Now, 
  \begin{equation}\begin{aligned}
  \label{eq}
  P[i \to i+1] = &P\text{[selecting a node $x$ with opinion $0$]}\times \\
  &P\text{[selecting a $y$, in the neighbourhood of the node $x$,} \\
  &\hspace{10pt}\text{with opinion $1$],} \\ 
  &\text{and,}\\
%   P[i \to i-1] = &P\text{[selecting a node $x$ with opinion $1$]}\times \\
  &P\text{[selecting a $y$, in the neighbourhood of the node $x$,}\\
  &\hspace{10pt}\text{with opinion $0$]}.
  \nonumber
  \end{aligned}
  \end{equation}
  \fi
  
  Therefore, we have,
  \begin{equation}\begin{aligned}
  \label{tp}
  P[i \to i+1] &= \sum_{\substack{x \in S(G), \\ s_x = 0}} \left( \frac{1}{N}   \sum\limits_{y \in S(G)}A_{xy}\frac{s_y}{k}\right) \\
  P[i \to i-1] &= \sum_{\substack{x \in S(G), \\ s_x = 1}} \left( \frac{1}{N}   \sum\limits_{y \in S(G)}A_{xy}\frac{1-s_y}{k}\right)
  \end{aligned}
  \end{equation}  
  
  Using a mean-field approximation, in the spirit of that used by Sood et al \cite{Sood2005}, the transition probabilities in equation \eqref{tp} become,
  \begin{equation}\begin{aligned}
  \label{tps}
  P[i \to i+1] &\approx \left(1-\frac{i}{N}\right)\left(\frac{i}{N}\right) \\
  P[i \to i-1] &\approx \left(\frac{i}{N}\right)\left(1-\frac{i}{N}\right).
  \end{aligned}
  \end{equation}

  For large $N$, the scaled process corresponding to the update rule $1$, $X^N[N^2t]/N$, converges to a diffusion $X(t)$ whose drift and diffusion terms derived using the transition probabilites from equation \eqref{tps} are,
    
  \begin{equation}
  \begin{aligned}
  \mu(x) &= 0 \\
  \sigma^2(x) &= 2x(1-x).
  \end{aligned}
  \end{equation}
  
  Additionally, if we define $\lambda \equiv N^2p$ (scaled jump), then for a small enough $p$ and a large enough $N$, the JVM (update rules 1 and 2) can be approximated by a superposition of the diffusion 
  derived above and a compound Poisson process, i.e., a jump diffusion process given as,
  
  \begin{equation}
  dX(t) = \sqrt{2X(t)(1-X(t))}dW(t) + Y dN(t)
  \label{jvmd}
  \end{equation}

  where $W(t)$ represents a Wiener process, and $N(t)$ represents a rate $\lambda$ Poisson process. We denote by $g(x)$ the probability density function of the jump random variable $Y$, and also note that $g(x)$ will have support $[-1, 1]$.
  The generator of this process is the integro-differential operator $\mathscr{L}$, defined by 
  
  \begin{equation} \label{ugen}
  \mathscr{L}f(x) = x(1-x)f''(x) + \lambda\int_{-\infty}^{+\infty}[f(x-y)-f(x)]g(y)dy. 
  \end{equation}
  
  Note here that as a result of using the mean field approximation, the jump voter diffusion given by equation \eqref{jvmd} does not have a term containing the degree parameter $k$.

  \subsection{Jump Threshold Voter Model}
  The jump threshold voter model (JTVM) is a discrete-time process that is updated according to two rules. At each time step, one of the following occurs: \newline
  \begin{enumerate}
   \item With probability $(1-p)$, a single node is randomly selected. If the number of opposing opinions in neighbourhood of the selected node is greater than or equal to a threshold $\theta$, then the opinion of the 
  originally selected node is updated. \newline
  \item With probability $p$, either a random number of $0$ opinion nodes update their opinions to $1$ or a random number of $1$ opinion nodes update their opinions to $0$. If this number exceeds the number of nodes available to be updated, then the 
   all the available nodes are updated. A signed form of this random variable $(Z)$ follows the convention that it is negative in the case of the former update, and positive for the latter.\newline
  \end{enumerate}
  
  As with the JVM, the mean of $Z$ is zero herein. \newline
  
  Table \ref{pars} summarises all the parameters discussed in this section.
  
  \begin{table}[H]
  \centering
  \caption[All model parameters and their brief meanings]{Summary of the model parameters.}
  \label{pars}
  \begin{tabular}{@{}llr@{}} \toprule 
  Parameter & Brief Description  \\ \midrule 
  $N$ & Total population size.  \\ 
  $k$ & Degree in a regular graph.  \\ 
  $p$ & Jump probability at each time step.  \\ 
  $Z$ (or $Y$) & Random variable that gives the jump or the external influence.   \\  
  $\mathrm{v}$ & Jump variance, or strength of the external influence. \\
  $\theta$ & Threshold parameter in the JTVM. \\
  $\lambda$ & Jump rate in the jump voter diffusion.\\
  \bottomrule 
  \end{tabular}
  \end{table}
  
\section{Results}
\label{results}
If $X(t)$ is a jump diffusion with generator $\mathscr{L}$, It\^{o}'s formula for jump processes \cite{prot1990} implies that 
\begin{equation}
M(t) \equiv f(X(t)) - \int_{0}^{t}\mathscr{L}f(X(s))ds
\end{equation}
is a martingale for any $C^2$ function $f(x)$. Application of the Optional Stopping Theorem (OST) to this martingale results in boundary value problems for both fixation probability and expected value of the consensus time 
for the jump voter diffusion. Before that, we mathematically define the first hitting times of the two absorbing states ($0$ and $1$) and the consensus time ($\tau$)

\begin{equation*} \begin{aligned}
  T_0 &\equiv \inf\{t \geq 0: X(t) \in (-\infty, 0]\},\nonumber \\ 
  T_1 &\equiv \inf\{t \geq 0: X(t) \in [1, \infty)\}, \\
  \tau &\equiv \inf\{t \geq 0: X(t) \in (-\infty, 0] \cup [1,\infty)\}.  
  \end{aligned}
\end{equation*}

The fixation probability $u(x) = P(T_1 < T_0|X(0)=x)$ satisfies,
\begin{equation}\label{upbvp} \begin{aligned}
\mathscr{L}u(x) &= 0, \\
u(x) = 0, \hspace{3pt}\forall \hspace{1pt}x &\in (-\infty, 0], \\ 
u(x) = 1,\hspace{3pt} \forall \hspace{1pt} x &\in [1, \infty). 
\end{aligned}
\end{equation}
  
Similarly, expected value of the consensus time $v(x) = E[\tau|X(0)=x]$ satisfies, 

\begin{equation}\begin{aligned}\label{uctbvp}
\mathscr{L}v(x) &= -1, \\
v(x) = 0, \hspace{3pt}\forall \hspace{1pt}x &\in (-\infty, 0] \cup [1, \infty).
\end{aligned}
\end{equation}
      
  \subsection{Jump Voter Model}
  We begin by investigating the fixation probability and the expected consensus time for the jump voter diffusion. To determine the fixation probability, we use the generator of the jump voter diffusion from equation \eqref{ugen} 
  in the BVP \eqref{upbvp}, to obtain,
  
  \begin{equation} \label{ufpide} \begin{aligned}
  x(1-x)u''(x) + \lambda\int_{-\infty}^{+\infty}&u(x-y)g(y)dy - \lambda u(x) = 0 \\
  u(x) &= 0, \hspace{3pt}\forall \hspace{1pt}x \in (-\infty, 0] \\
  u(x) &= 1,\hspace{3pt} \forall \hspace{1pt} x \in [1, \infty).
  \end{aligned}
  \end{equation}
  
\begin{figure}[H]
  \centerline{\includegraphics[width=0.4\textwidth]{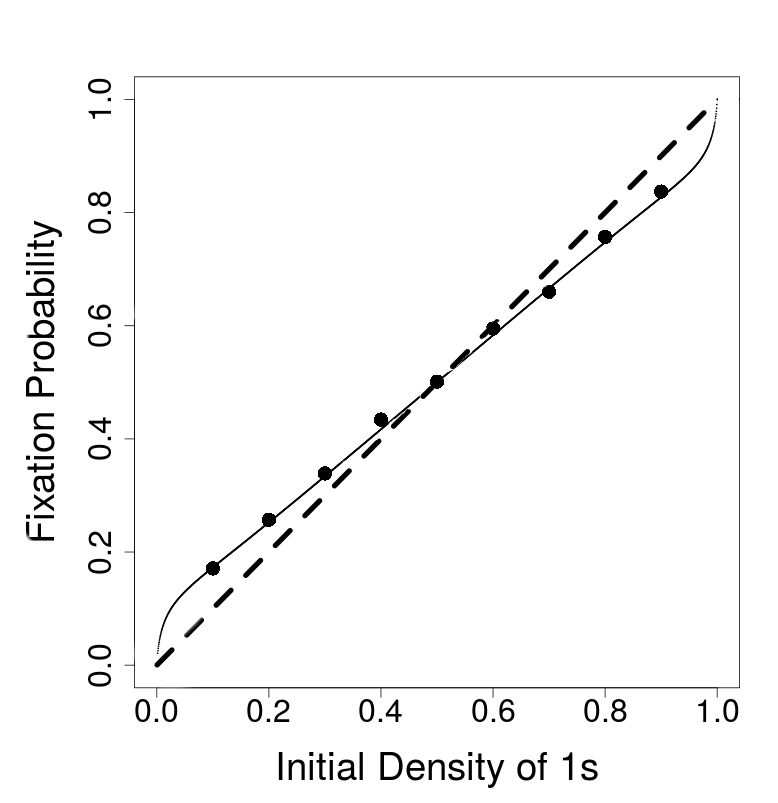}}
  \caption[Fixation probability for the jump voter model]{Fixation probability for the JVM on a regular graph with $N = 500, k = 4, p = 1/(500 \times 10), \mathrm{v} = 0.04$ (see Table \ref{pars} for meaning of parameters). 
  $(\boldmath - -)$ denotes the solution for the VM, $(\boldmath \textendash\textendash)$ denotes the numerical solution of equation \eqref{ufpide}, $(\bullet)$ denotes simulation results based on update rules in Section $2.1$, where each point is obtained by averaging
  over $1000$ runs.  The external influence, $Z$, has a truncated normal distribution.}
  \label{ufp}
  \end{figure}

  It is quite challenging to find an analytical solution for a variable-coefficient integro-differential equation such as equation \eqref{ufpide}. Even a similar constant-coefficient equation requires imposing some structure on the function $g(x)$ to 
  make a closed-form solution possible \cite{Kou2004}. We thus solve this problem numerically and compare the solution to Monte Carlo simulations of the JVM (Figure \ref{ufp}).
    
  Similarly, if we expand equation \eqref{uctbvp} using the generator in equation \eqref{ugen}, we obtain
  
  \begin{equation} \label{uctpide} \begin{aligned}
  x(1-x)v''(x) + \lambda&\int_{-\infty}^{+\infty}v(x-y)g(y)dy - \lambda v(x) = -1 \\
  v(x) =  0, \hspace{3pt}&\forall \hspace{1pt}x \in (-\infty, 0] \cup [1, \infty).
  \end{aligned}
  \end{equation}
  
\begin{figure}[H]
  \centerline{\includegraphics[width=0.4\textwidth]{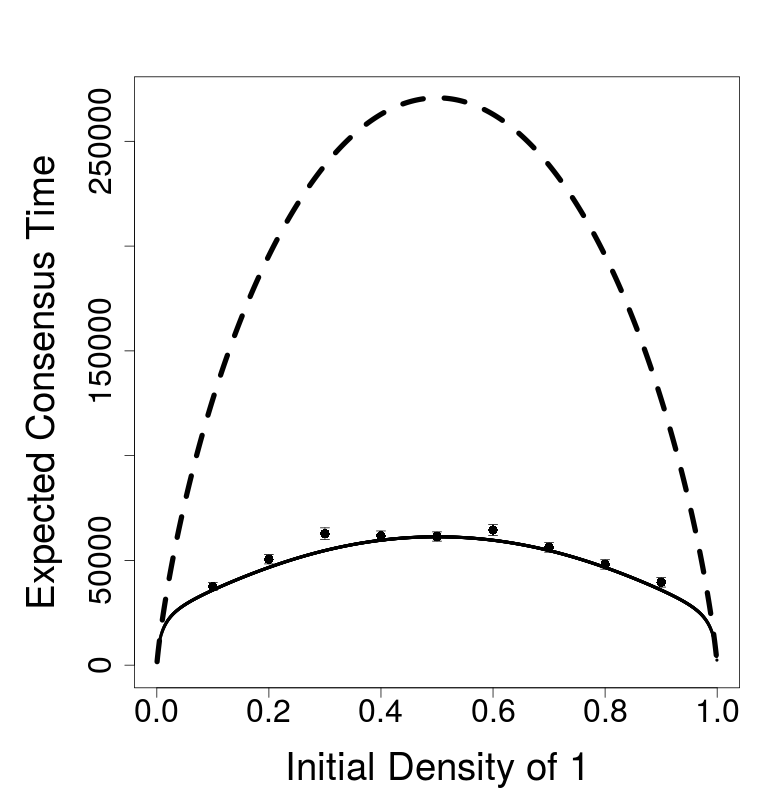}}
  \caption[Expected consensus time for the jump voter model]
  {Expected consensus time for the JVM on a $25 \times 25$ lattice, with $p = 1/(625 \times 10), \mathrm{v}=0.03$ (see Table \ref{pars} for meaning of parameters). 
  $(\boldmath - -)$ denotes the solution for the VM, $(\boldmath \textendash\textendash)$ denotes the numerical solution of equation \eqref{uctpide}, $(\bullet)$ denotes simulations results based on update rules in Section $2.1$, where each 
  point is obtained by averaging over $500$ runs, and error bars indicate standard error of the mean. The external influence, $Z$, has a truncated normal distribution.}
  \label{uct}
  \end{figure}

  Solving the problem numerically, we find that the solution again closely matches the results of the Monte Carlo simulations of the JVM (Figure \ref{uct}).
    
  We notice in Figure \ref{uct} that although our solution for the consensus time has properties that are qualitatively similar to those of the VM, the quantitative difference between the two is considerable. 
  The two parameters, jump probability $p$ and jump variance $\mathrm{v}$, together determine the overall impact of the external influence, and we collectively refer to them as jump parameters. We find that 
  even for fairly small jump parameter values, the expected consensus time differs dramatically between the VM and the JVM.
    
    \iffalse
    \begin{figure}[H]
    
    \centering
    
    \subfigure[]{\includegraphics[width=6cm]{fpcomp} \label{fpcomp}}
    \;
    \subfigure[]{\includegraphics[width=6cm]{ctcomp} \label{ctcomp}}
    
    \caption[Comparison between the voter model and the jump voter model]
    {(a) Fixation probability and (b) expected consensus time comparison between the voter model and the jump voter model. $N=500$ and for the jump voter model, $p=1/(500 \times 10), \mathrm{v}=0.03$ (see Table \ref{pars} for meaning of 
    parameters). This comparison corresponds to comparing model A and model D (or model B and model E), as shown in Table \ref{chart}.}
    \label{comp}
    \end{figure}
    \fi

\begin{figure}[H]
    
    \centering
    
    \subfigure[]{\includegraphics[width=6cm]{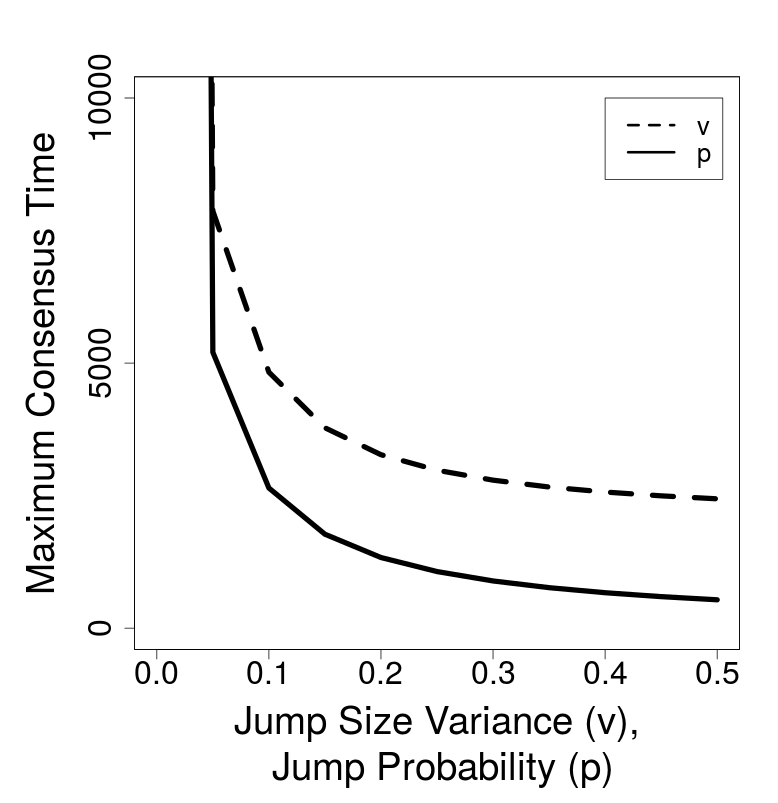} \label{ctvsvp}}
    \;
    \subfigure[]{\includegraphics[width=6cm]{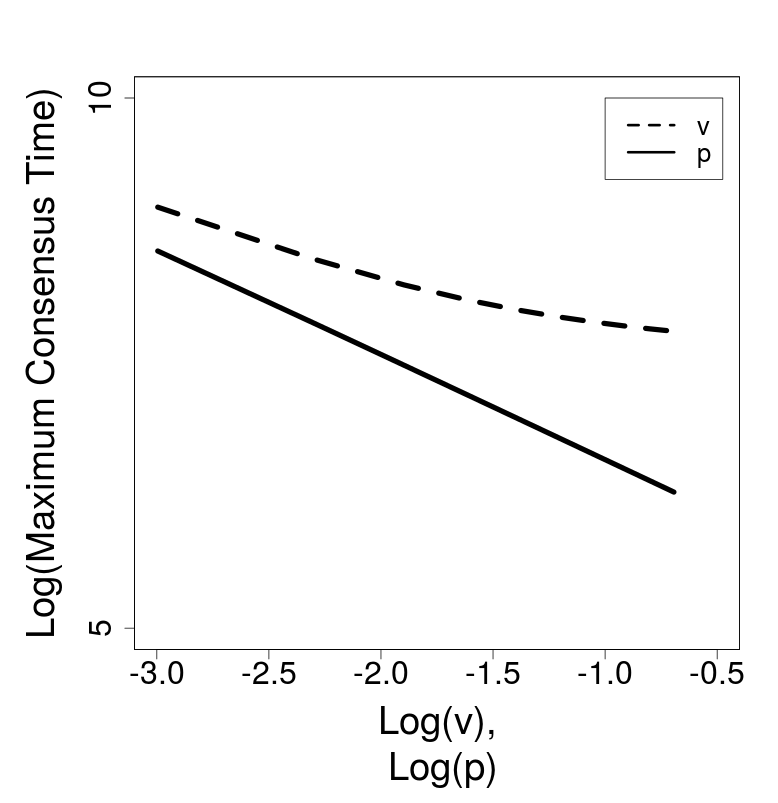} \label{ctvsvpl}}
    
    \caption[Expected consensus time dependence on jump parameters for the jump voter model - part 2]
    {(a) Maximum consensus time for the jump voter diffusion as a function of the jump parameters individually, $N=500$. (The maximum consensus time axis has been restricted to $10000$ time steps for diagrammatic clarity.) The maximum consensus time decreases rapidly
    as $p$ and $\mathrm{v}$ increase. For plotting the dependence on $p$, $\mathrm{v} = 0.001$, and for plotting the dependence on $\mathrm{v}$, $p=0.001$ (see Table \ref{pars} for meaning of parameters). Plot (b) is plot (a) on a log-log scale.}
    \label{ctvsvandp}
    \end{figure}

    If we consider just the maximum value of consensus time (which occurs at initial density of $1$s is $0.5$), we can plot it as a function of the two jump parameters (Figure \ref{ctvsvandp}). We see that the consensus time decreases 
    rapidly as a function of $p$ and $\mathrm{v}$, especially at small values.

    \iffalse
    \begin{figure}[H]
    \centerline{\includegraphics[width=0.4\textwidth]{ctvsvp.png}}
    \caption[Expected consensus time dependence on jump parameters for the jump voter model - part 1]
    {Maximum consensus time for the jump voter diffusion as varying with the two jump parameters, $N=500$. (The maximum consensus time axis has been restricted to $100$ time steps for diagrammatic clarity.)}
    \label{ctvsvp}
    \end{figure}
    
    We also isolate the dependence of the consensus time on the individual jump parameters. The surface in Figure \ref{ctvsvp} appears symmetric about $p=\mathrm{v}$. However, examination of two curves on the surface (Figure \ref{ctvsvandp}) indicates
    that the symmetry is only approximate. The parameter $p$ has a slightly stronger effect on consensus time than parameter $\mathrm{v}$.
    \fi
    
    \textit{Consensus time in the VM is sensitive to the presence of jumps.} 
    Consensus time decreases in both jump parameters ($p$ and $\mathrm{v}$), and the rate of decrease of consensus time is very high at low jump parameter values, and significantly drops as parameter
    values increase (Figure \ref{ctvsvp}). Therefore, it is primarily the presence of jumps that appears to be a key factor for the consensus time.
    Consistent with that observation, the dependence of consensus time on both jump parameters exhibits a power-law like nature (Figure \ref{ctvsvpl}).    
    Overall, jumps expedite consensus, introducing little skew in addition to that inherently present due to the initial densities. Therefore, jumps may be an important ingredient to consider in opinion dynamics models based on the VM.
    
      \subsection{Jump Threshold Voter Model}
\begin{figure}[H]
    
    \centering
    
    \subfigure[]{\includegraphics[width=6cm]{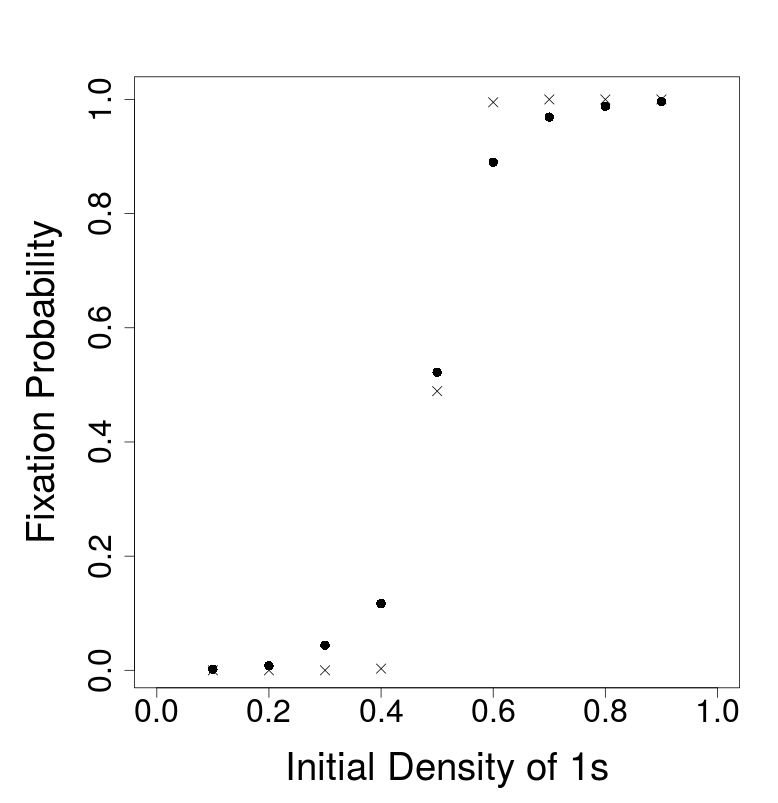} \label{fpjtvm}}
    \;
    \subfigure[]{\includegraphics[width=6cm]{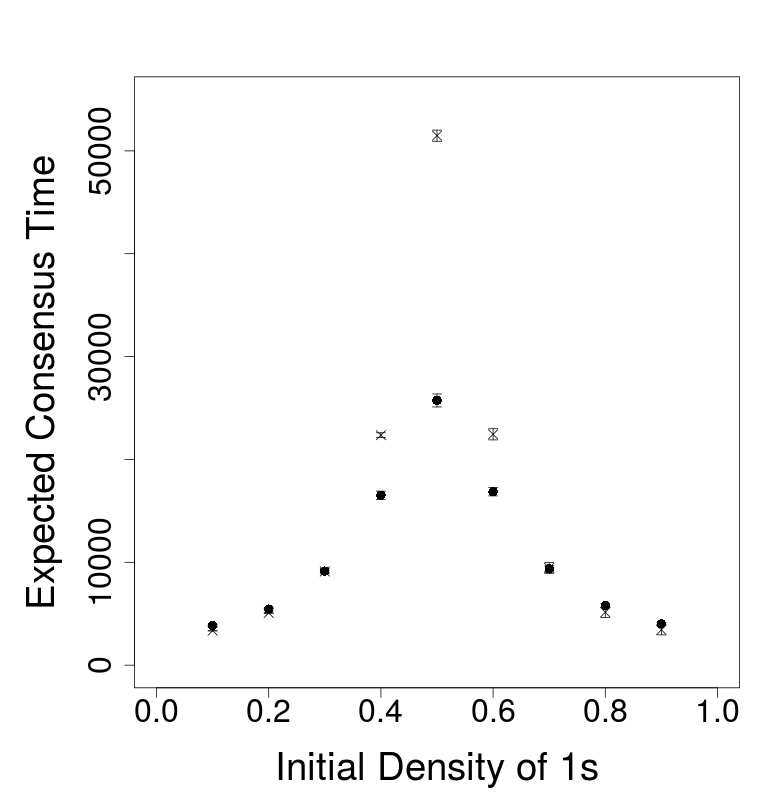} \label{ctjtvm}}
    
    \caption[Fixation probability and expected consensus time for the jump threshold voter model]
    {(a) Fixation probability and (b) expected consensus time for the JTVM $(\bullet)$ in comparison with the TVM $(\times)$, on a $25 \times 25$ lattice with $p = 1/(625\times 10), \mathrm{v} = 0.03, \theta=2$ 
    (see Table \ref{pars} for meaning of parameters). Each point is based on $1000$ runs, and error bars indicate standard error of the mean. The external influence, $Z$, has a truncated normal distribution.
    }
    \label{jtvmres}
    \end{figure}

      In this section, we study the fixation probability and consensus time for the JTVM. The results are obtained through Monte Carlo simulations, and are shown in Figure \ref{jtvmres}.
  
\begin{figure}[H]
  \centering
  \subfigure[JTVM]
  {
  \includegraphics[scale=0.3]{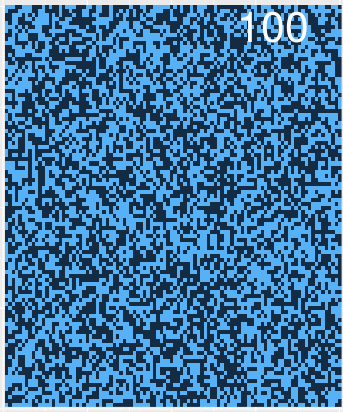}
  \includegraphics[scale=0.3]{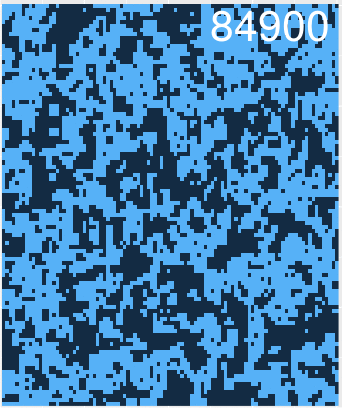}
  \includegraphics[scale=0.3]{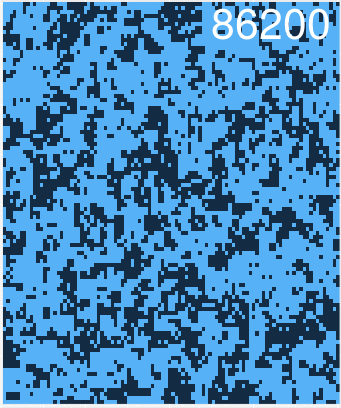}
  \includegraphics[scale=0.3]{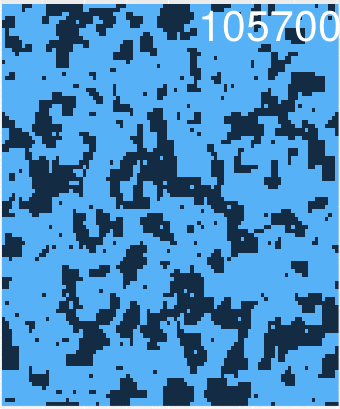}
  \includegraphics[scale=0.3]{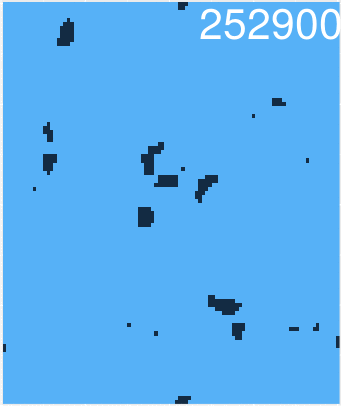}
  \label{jtvmsim}
  }
  
  \subfigure[JVM]
  {
  \includegraphics[scale=0.3]{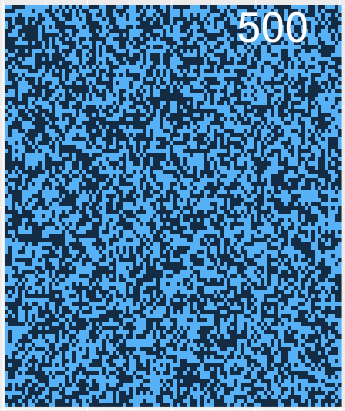}
  \includegraphics[scale=0.3]{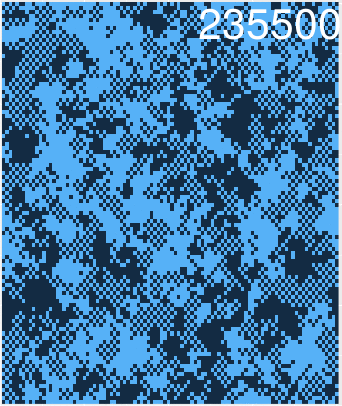}
  \includegraphics[scale=0.3]{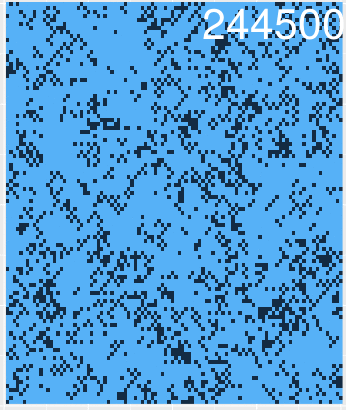}
  \includegraphics[scale=0.3]{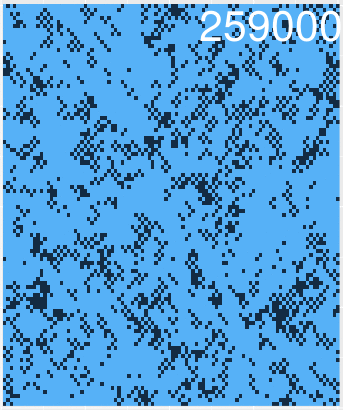}
  \includegraphics[scale=0.3]{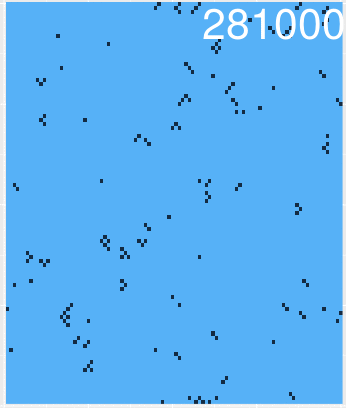}
  \label{jvmsim}
  }
  \caption[Spatial comparison between the jump voter model and the jump threshold voter model]
  {Snapshots of evolution of (a) the JTVM, and (b) the JVM on a $100 \times 100$ lattice, with $p = 1/(10000 \times 5), \mathrm{v} = 0.03, \theta=2$
  (see Table \ref{pars} for meaning of parameters). Initial density of $1$s is $0.5$ for both models. The numbers in the top right corner denote the time step for the respective panel. The external influence, $Z$, has a truncated normal distribution.
  }
  \label{simcomp}
  \end{figure}

  The spatial structure of the dynamics yields further insight into the model behaviour. Typical spatial results are shown in Figure \ref{simcomp}.
  It is known that the evolution of clusters in the TVM is characterised by motion by mean curvature \cite{castellano2009, dall2007}. The introduction of jumps to this model plays the role 
  of disrupting the clustering sporadically. This can be seen in Figure \ref{jtvmsim}.  
  
  \textit{The JTVM is similar, for the fixation probability and the consensus time, to the TVM at low initial minority densities.} The differences between the TVM and the JTVM become prominent only at initial minority densities greater 
  than $0.4$ approximately, which we call the ``critical'' density. The fixation probability for the TVM exhibits behaviour similar to a step function (Figure \ref{jtvmres}), where consensus is almost always reached on 
  the opinion in majority at the start of the process. The TVM thus amplifies the advantage held by to a particular opinion type due to a higher initial density. Comparing the fixation probability in this case to that for the  
  JTVM, we notice that the probabilities become less extreme at initial minority densities greater than the critical density. Therefore, at initial minority densities greater than the critical density, jumps help moderate the advantage 
  amplification inherent to the TVM. Next we assess the effect of jumps on the consensus time. Jumps have an effect of increasing fluctuations in the opinion density. This effect appears more pronounced at initial minority densities 
  higher than the critical density, as we obtain a quicker consensus in that regime. 
  
  Overall, for initial minority densities lower than the critical density, the clustering effect is too strong to be disrupted by jumps. However, once the initial minority density exceeds a certain value, the disrupting effects of jumps begin to counter 
  the clustering effect inherent to the TVM. 
  
  \textit{Jumps expedite consensus in the TVM less than the VM.} We notice, based on comparing Figures\ref{uct} 
  and \ref{jtvmres}, that jumps reduce the consensus time significantly more in the VM than in the TVM. This suggests that the TVM is more robust to an external influence. This 
  is again attributable to motion by mean curvature in the TVM. Consider the extreme case, for both the JVM and JTVM, where the jump causes a $0 \to 1$ flip at a node in the interior of 
  a $0$ cluster. In the JVM, there is a non-zero probability that the $1$ in the interior can in turn cause a $0 \to 1$ flip at one of its neighbouring nodes. But in the TVM, the probability that this opinion $1$ node 
  can flip any of its opinion $0$ neighbours is zero. Therefore, the cluster patterns of the VM tend to be more vulnerable to jumps as compared to those of the TVM. This resistance serves as a possible explanation for 
  the higher robustness of the TVM to jumps with regards to the consensus time. This phenomenon can also be observed in Figure \ref{simcomp}: Notice the similarity in the cluster pattern in the second and fourth panels of 
  Figure \ref{jtvmsim} despite the occurence of a jump in the third panel. On the contrary, no such trend is seen in Figure \ref{jvmsim}.

\section{Discussion}
\label{disc}
In this work, we have developed extensions of the voter model (VM) and the threshold voter model (TVM), that incorporate an external influence in the form of many opinions shifting in the same direction simultaneously.
Most existing literature on opinion dynamics only studies opinion evolution under influences internal to the system. This work provides a systematic study of opinion dynamics under an additional external influence. 
We approximated the jump voter model (JVM) by means of a jump diffusion process. This approach allowed us to analytically determine the
probability of reaching consensus on opinion $1$ (fixation probability) and the mean consensus time. For the jump threshold voter model (JTVM), we relied on simulations to determine fixation probability 
and mean consensus time.

The diffusion approximation for the VM does not depend on the degree of the regular graph \cite{Sood2005}, and we note that this property is retained in the JVM.
This means that the fixation probability and consensus time results for the JVM will remain true for all regular graphs ranging from a complete graph (where $k=N-1$) to even a cycle graph (where $k=2$). Thus, in a society where everyone 
has the same neighbourhood size, fixation probability and consensus time are independent of the neighbourhood size as long as agents update their opinions by a combination of random sampling from their neighbourhood and by an unbiased external influence.
  
Another key observation is that jumps expedite consensus more in the VM than in the TVM. Thus, in a society where agents update 
their opinion if the pressure from their neighbourhood is sufficient (based on a threshold parameter), external influence has a lesser effect on consensus time than in a society where agents update their opinions by randomly sampling from 
their neighbourhood. 
  
This work opens up multiple interesting directions that may be further explored. The domain of network science is currently expanding very rapidly, and one natural extension of our work is to study our models on a heterogeneous graph structure 
such as a scale-free network. Such work could lead to pragmatic insights since scale-free networks have been shown to be ubiquitous in various real-world social systems \cite{barabasi1999}.
    
\bibliographystyle{siam}
\bibliography{bibliography}

\begin{thebibliography}{10}

\bibitem{asch1956}
{\sc S.~E. Asch}, {\em Studies of independence and conformity: I. a minority of
  one against a unanimous majority.}, Psychological monographs: General and
  applied, 70 (1956), p.~1.

\bibitem{barabasi1999}
{\sc A.-L. Barab{\'a}si and R.~Albert}, {\em Emergence of scaling in random
  networks}, science, 286 (1999), pp.~509--512.

\bibitem{Baronchelli2005}
{\sc A.~Baronchelli, M.~Felici, E.~Caglioti, V.~Loreto, and L.~Steels}, {\em
  {Sharp transition towards shared vocabularies in multi-agent systems}}, 06014
  (2005), p.~12.

\bibitem{beck2002}
{\sc P.~A. Beck, R.~J. Dalton, S.~Greene, and R.~Huckfeldt}, {\em The social
  calculus of voting: Interpersonal, media, and organizational influences on
  presidential choices}, American Political Science Review, 96 (2002),
  pp.~57--73.

\bibitem{ben1996}
{\sc E.~Ben-Naim, L.~Frachebourg, and P.~Krapivsky}, {\em Coarsening and
  persistence in the voter model}, Physical Review E, 53 (1996), p.~3078.

\bibitem{Brulle2012}
{\sc R.~J. Brulle, J.~Carmichael, and J.~C. Jenkins}, {\em {Shifting public
  opinion on climate change: an empirical assessment of factors influencing
  concern over climate change in the U.S., 2002–2010}}, Climatic Change, 114
  (2012), pp.~169--188.

\bibitem{castellano2009}
{\sc C.~Castellano, S.~Fortunato, and V.~Loreto}, {\em Statistical physics of
  social dynamics}, Reviews of modern physics, 81 (2009), p.~591.

\bibitem{Clifford1973}
{\sc P.~Clifford and A.~Sudbury}, {\em {A Model for Spatial Conflict}},
  Biometrika, 60 (1973), p.~581.

\bibitem{Cox1991}
{\sc J.~T. Cox and R.~Durrett}, {\em {Nonlinear Voter Models}}, Random Walks,
  Brownian Motion, and Interacting Particle Systems,  (1991), pp.~189--202.

\bibitem{DallAsta2006}
{\sc L.~Dall'Asta, A.~Baronchelli, A.~Barrat, and V.~Loreto}, {\em
  {Nonequilibrium dynamics of language games on complex networks.}}, Physical
  review. E, Statistical, nonlinear, and soft matter physics, 74 (2006),
  p.~036105.

\bibitem{dall2007}
{\sc L.~Dall'Asta and C.~Castellano}, {\em Effective surface-tension in the
  noise-reduced voter model}, EPL (Europhysics Letters), 77 (2007), p.~60005.

\bibitem{DeGroot1974}
{\sc M.~H. DeGroot}, {\em {Reaching a Consensus}}, Journal of the American
  Statistical Association, 69 (1974), pp.~118--121.

\bibitem{druckman2005}
{\sc J.~N. Druckman}, {\em Media matter: How newspapers and television news
  cover campaigns and influence voters}, Political Communication, 22 (2005),
  pp.~463--481.

\bibitem{Fotouhi2013}
{\sc B.~Fotouhi and M.~G. Rabbat}, {\em {The Effect of Exogenous Inputs and
  Defiant Agents on Opinion Dynamics With Local and Global Interactions}}, IEEE
  Journal of Selected Topics in Signal Processing, 7 (2013), pp.~347--357.

\bibitem{Hegselmann2002}
{\sc R.~Hegselmann and U.~Krause}, {\em {Opinion Dynamics and Bounded
  Confidence}}, Simulation, 5 (2002), p.~2.

\bibitem{HolleyRichardandLiggett1975}
{\sc R.~Holley and T.~Liggett}, {\em {Ergodic theorems for weakly interacting
  infinite systems and the voter model}}, The annals of probability, 3 (1975),
  pp.~643--663.

\bibitem{keynes2006}
{\sc J.~M. Keynes}, {\em General theory of employment, interest and money},
  Atlantic Publishers \& Dist, 2006.

\bibitem{Kou2007}
{\sc S.~G. Kou}, {\em {Chapter 2 Jump-Diffusion Models for Asset Pricing in
  Financial Engineering}}, Handbooks in Operations Research and Management
  Science, 15 (2007), pp.~73--116.

\bibitem{Kou2004}
{\sc S.~G. Kou and H.~Wang}, {\em {Option Pricing Under a Double Exponential
  Jump Diffusion Model}}, Management Science, 50 (2004), pp.~1178--1192.

\bibitem{le1897}
{\sc G.~Le~Bon}, {\em The crowd: A study of the popular mind}, Fischer, 1897.

\bibitem{VanZandt}
{\sc T.~Liggett}, {\em {Coexistence in Threshold Voter Model}}, The Annals of
  Probability, 22 (1994), pp.~764--802.

\bibitem{mackay}
{\sc C.~Mackay}, {\em Extraordinary popular delusions and the madness of
  crowds}, Start Publishing LLC, 2012.

\bibitem{milgram1974}
{\sc S.~Milgram and R.~Fleissner}, {\em Das Milgram-Experiment}, Rowohlt
  Reinbek, 1974.

\bibitem{nyczka2012}
{\sc P.~Nyczka, K.~Sznajd-Weron, and J.~Cis{\l}o}, {\em Phase transitions in
  the q-voter model with two types of stochastic driving}, Physical Review E,
  86 (2012), p.~011105.

\bibitem{prot1990}
{\sc P.~Philip}, {\em Stochastic integration and differential equations}, 1990.

\bibitem{Quattrociocchi2014}
{\sc W.~Quattrociocchi, G.~Caldarelli, and A.~Scala}, {\em {Opinion dynamics on
  interacting networks: media competition and social influence.}}, Scientific
  reports, 4 (2014), p.~4938.

\bibitem{schelling1971}
{\sc T.~C. Schelling}, {\em Dynamic models of segregation†}, Journal of
  mathematical sociology, 1 (1971), pp.~143--186.

\bibitem{sherif1936}
{\sc M.~Sherif}, {\em The psychology of social norms.},  (1936).

\bibitem{Sood2005}
{\sc V.~Sood and S.~Redner}, {\em {Voter model on heterogeneous graphs}},
  Physical Review Letters, 94 (2005), pp.~6--9.

\bibitem{vazquez2008}
{\sc F.~Vazquez et~al.}, {\em Analytical solution of the voter model on
  uncorrelated networks}, New Journal of Physics, 10 (2008), p.~063011.

\end{thebibliography}
\end{document}